\documentclass[fleqn,twoside]{article}
\usepackage{espcrc2}
\usepackage{graphicx}
\usepackage{bm}

\newcommand{\AmS}{{\protect\the\textfont2
  A\kern-.1667em\lower.5ex\hbox{M}\kern-.125emS}}

\title{Confinement from Instantons or Merons\thanks{Supported by DOE cooperative ageement DE-FC02-94ER40818.}}

\author{J. W. Negele,\address{Center for Theoretical Physics, 
Massachusetts Institute of Technology,  Cambridge,
MA 02139, USA}
F. Lenz,\address{Institute for Theoretical Physics III,
University of Erlangen-N\"urnberg,
 Erlangen, Germany}
and
M. Thies$^{\rm b}$}
       
\begin{document}

\begin{abstract}
In contrast to ensembles of singular gauge instantons, which are well known to fail to produce confinement, it is shown that effective theories based on ensembles of  merons or regular gauge instantons  do produce confinement. Furthermore, when the scale is set by the string tension, the action density, topological susceptibility, and glueball masses are similar to those arising in lattice QCD.
\end{abstract}
\maketitle
\section{INTRODUCTION}
\label{sec:introduction}

This work brings new insight to the venerable question of whether there is a low energy effective theory based on  classical (pseudoparticle) degrees of freedom that embodies the two essential features of SU(2) Yang Mills theory: confinement and chiral symmetry breaking. We first  consider all the relevant gauge field degrees of freedom, 
$a_\mu(z_i,h_i)$, associated with a pseudoparticle at position $z_i$ with color orientation $h_i$. We then construct  gauge fields  from an ensemble of such pseudoparticles, $A_\mu = \sum_i a_\mu(z_i,h_i)$, and calculate the observable properties of the effective theory by evaluating the partition function,
\begin{equation}
\langle {\cal O}\rangle=\frac{1}{Z} \int dz_i dh_i e^{-\frac{1}{g^2} S[A{z_i,h_i} ] }
{\cal O}[A{z_i,h_i}],
\end{equation}
using Metropolis sampling of the parameters ${\{z_i,h_i \}}$. Finally, for effective theories 
yielding confinement, we compare the action density, $\langle s \rangle$, topological susceptibility, $\chi$, and glueball masses expressed in units of the string tension,  $\sigma$,  with lattice QCD.

Based on the idea that classical tunneling solutions between topological sectors and their fluctuations should play a dominant role in the QCD path integral, 
as in ref.~\cite{Lenz:2003jp},  we consider 3 classes of classical pseudoparticle solutions to SU(2) gauge theory:  singular gauge instantons with 
\hbox{$a_\mu^a=\frac{\rho^2}{x^2} \frac{2 \eta_{a \mu \nu} x_\nu}{x^2 + \rho^2}$,} 
regular gauge instantons with 
\hbox{$a_\mu^a= \frac{2 \eta_{a \mu \nu} x_\nu}{x^2 + \rho^2}$,}
and merons with 
\hbox{$a_\mu^a=\frac{ \eta_{a \mu \nu} x_\nu}{x^2 + \rho^2}$.} The effective theory in each case is calculated analogously by superposing pseudoparticle solutions and using Metropolis updating of the coordinates and color orientations to sample the action.

For singular gauge instantons, the effective theory corresponds to the highly successful instanton model reviewed in ref.\cite{Schafer:1996wv}, which  incorporates chiral symmetry breaking and many physical aspects of low energy QCD, but fails to produce confinement. Since the gauge fields fall off $\sim\!\!\frac{1}{x^3}$, fields from adjacent instantons have little overlap and Metropolis updating induces negligible color orientations. Then, as in the analytic calculation of the static potential in a dilute instantion gas\cite{Callan:1978ye}, when the separation between two Polyakov lines is larger than the size of an instanton, each instanton affects only one line, and  integration over all color orientations eliminates any contribution to a linear potential.

Regular gauge instanton gauge fields fall off $\sim\!\!\frac{1}{x}$, so that in general, superposition yields fields with large average fields between instantons rendering them energetically unfavorable. The dominant configurations selected by Metropolis sampling contain correlations in color orientation that render the average field small while retaining lumps of $FF$ and $\tilde F F$ at the positions of instantons and antiinstantons. Repeating the confinement argument for two Polyakov lines, an instanton interacting with one line cannot have a color orientation independent of an instanton interacting with the other, so there is no reason for the confining potential to be eliminated.

As argued in ref~\cite{Lenz:2003jp}, we also consider merons. Although individual merons have infrared logarithmic singularities in the action,  these singularities can easily be cancelled for sets of multiple merons and ensembles of merons  behave essentially the same as regular gauge instantons.

		\begin{figure}[t]
		 \vspace*{-1.1cm}
	  \begin{tabular}{*{3}{c}}
	 	\hspace*{-0.8cm}  
		\includegraphics[clip=true,angle=-90,scale=0.3]{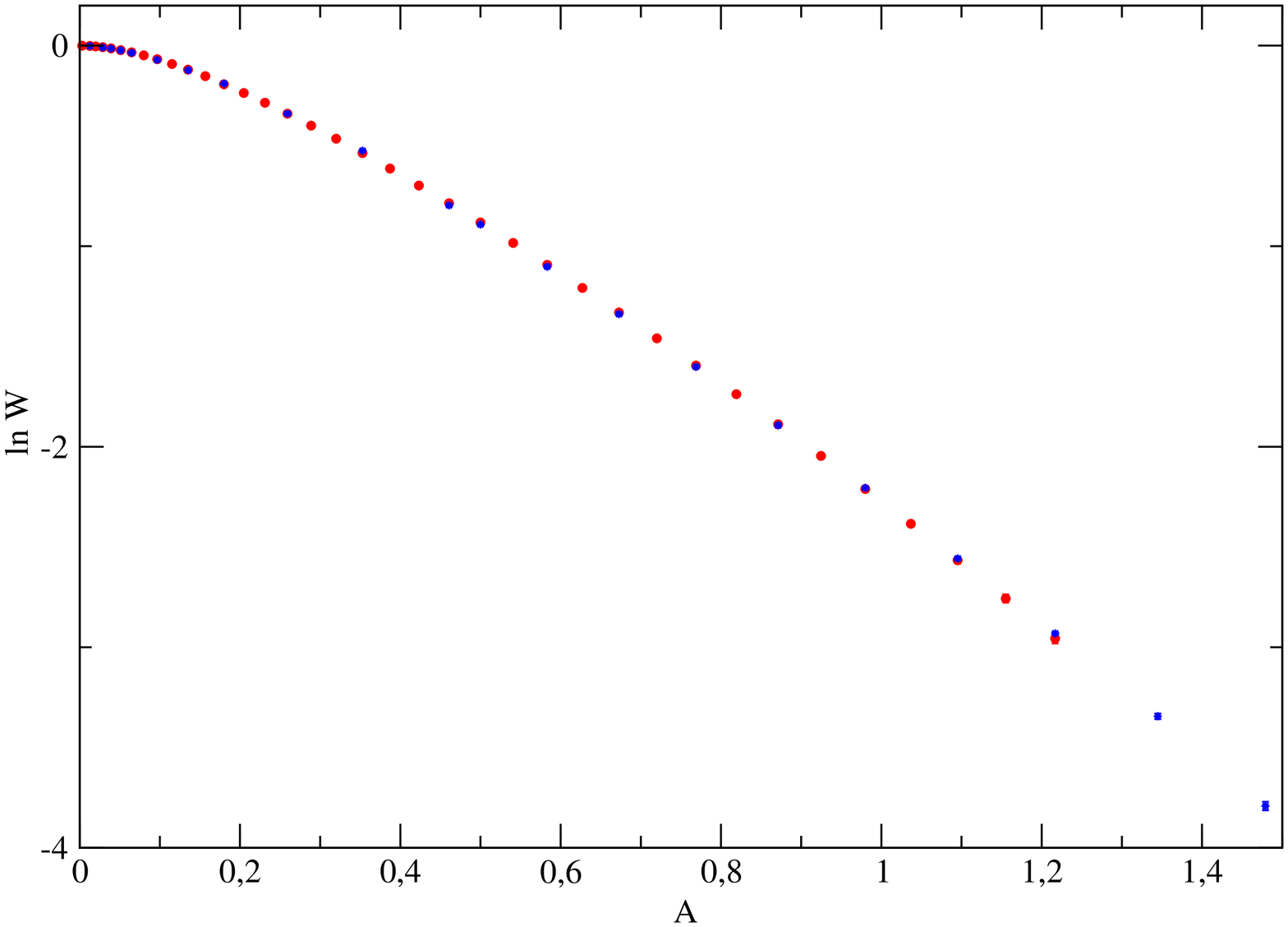} \\[-1.2cm]
		\hspace*{-0.8cm}  
		   \includegraphics[clip=true,angle=-90,scale=0.3]{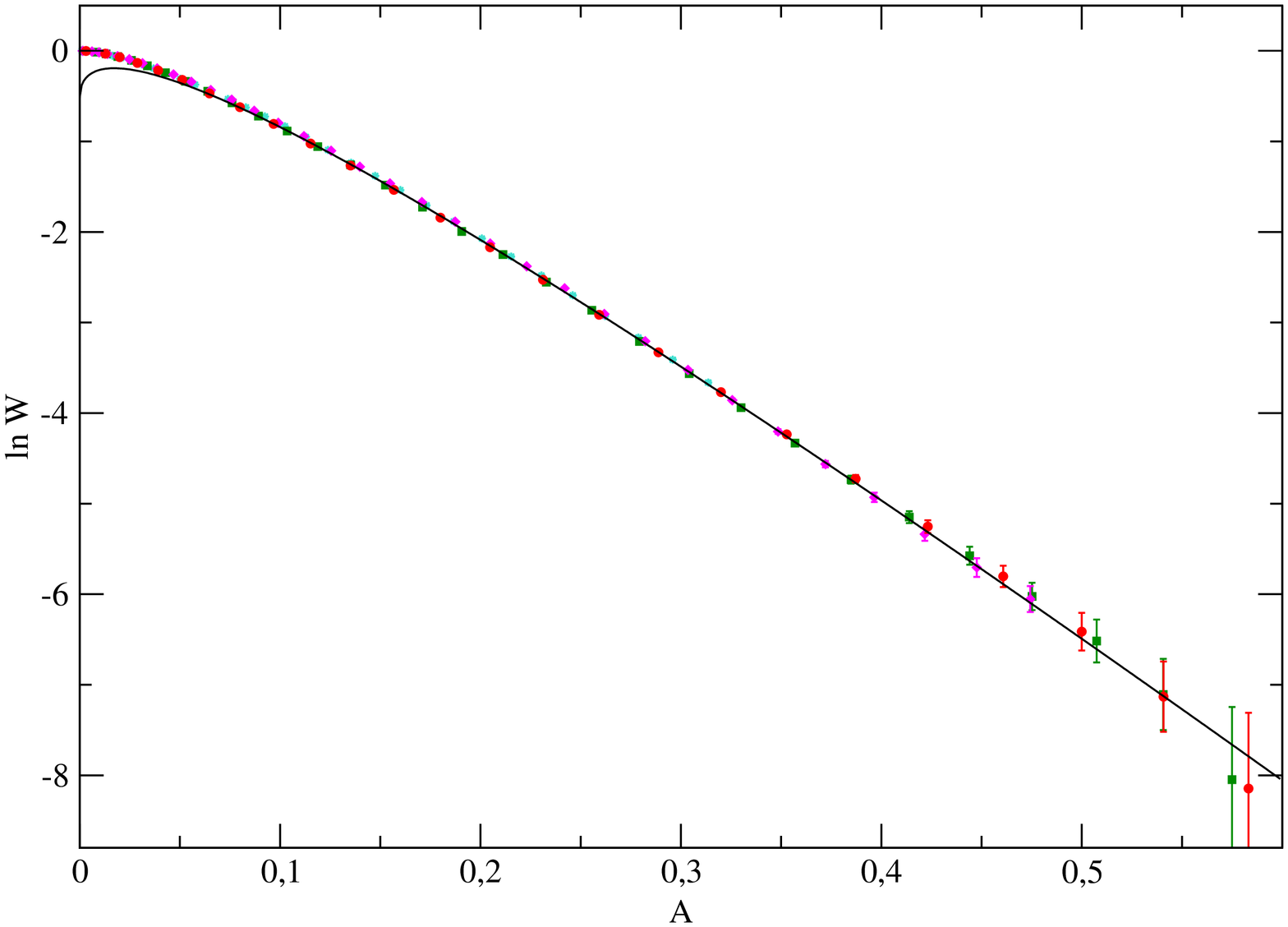} \\
 	   \end{tabular}		   		   
	  \vspace*{-1.5cm}
		  \caption{Logarithm of Wilson loops as a function of  area for ensembles of merons (top) and regular gauge instantons (bottom).  The universal curve for 50 merons in a $2^4$ box and 800 merons in a $4^4$ box confirms the thermodynamic limit. The universal curve
for 50, 100, 200, and 500 instantons demonstrates the scaling described in the text.}
		  \label{fig:wilson}
		  \vspace*{-0.5 cm}
		\end{figure}
		
\section{CONFINEMENT}

As shown in Fig.~\ref{fig:wilson}, the most striking result is that both regular gauge instantons and merons produce Wilson loops with area law behavior indicating confinement. An important technical issue with long range gauge fields is ensuring a correct continuum limit, and the top graph shows that ensembles of merons calculated on two volumes differing by a factor of 16 are indistinguishable. 

The lower graph in Fig.~\ref{fig:wilson} displays universal scaling of instanton Wilson loops, which also occurs for merons.  In lattice QCD, which depends on lattice spacing $a$ and $g^2$, it is convenient to set the scale by the string tension $\sigma$, so one may pick $g^2$ and use $\sigma$ to determine $a$ in physical units. In the instanton effective theory, there are three parameters, the size $\rho$, $g^2$, and density of instantons $n_I$, so we may pick $g^2$ and use $\sigma$, to determine one combination of $\rho$ and $n_I^{-1/4}  $in physical units.  Remarkably, the observables of interest are  insensitive to the other combination. Operationally, for each set \{$\rho, n_I, g^2$\}, we rescale the area $A \to \lambda A $ and size $\rho \to \sqrt{\lambda} \rho$ and adjust the $\lambda$'s to obtain the universal curve as shown in Fig.~\ref{fig:wilson}. Scaling the four sets of data in Fig.~\ref{fig:wilson} with instanton number changing by a factor 10 at fixed $\rho$, $g^2$, and volume and using $\sqrt{\sigma} = 410$ MeV yields the results for the action density $\langle s \rangle$ and topological susceptibility, $\chi$, shown in Table~\ref{tab:scaled}.  These results agree qualitatively with lattice data from refs.~\cite{Campostrini:1983nr,DiGiacomo:1989id,Lucini:2001ej}, and similar agreement is obtained with merons.

\begin{table}[t]
\begin{center}
	\caption{Action density and topological susceptibility for regular gauge instanton ensembles compared with lattice QCD.}
	\vspace{.3cm}
\begin{tabular}{|c|c|c|c|}  \hline 
$n_{I}$  &$\rho$& $\langle  s\rangle\;\;$  & $\chi^{1/4}$ \\  \hline
[fm $^{-4}]$ &[fm]&[fm $^{-4}]$  &[MeV]   
\\ \hline \hline
1.68  &0.33& $291 $ &162 \\ \hline
1.54&$ 0.27$ & $307 $  & 164\\ \hline
1.45& $ 0.23$ & $ 314$  &180\\ \hline
1.64& $ 0.19 $ & $340$& 190\\ \hline 
lattice & & 80-446 & 198 $\pm$ 3 \\ \hline
\end{tabular}
\end{center}
\vspace{-0.8cm}
\label{tab:scaled}
\end{table}%

\section{CORRELATION FUNCTIONS}

		\begin{figure}[!t]
		  \vspace*{-1.1cm}
		 \begin{tabular}{*{3}{c}}
	 	\hspace*{-0.8cm} 
		  \includegraphics[clip=true,angle=-90,scale=0.3]{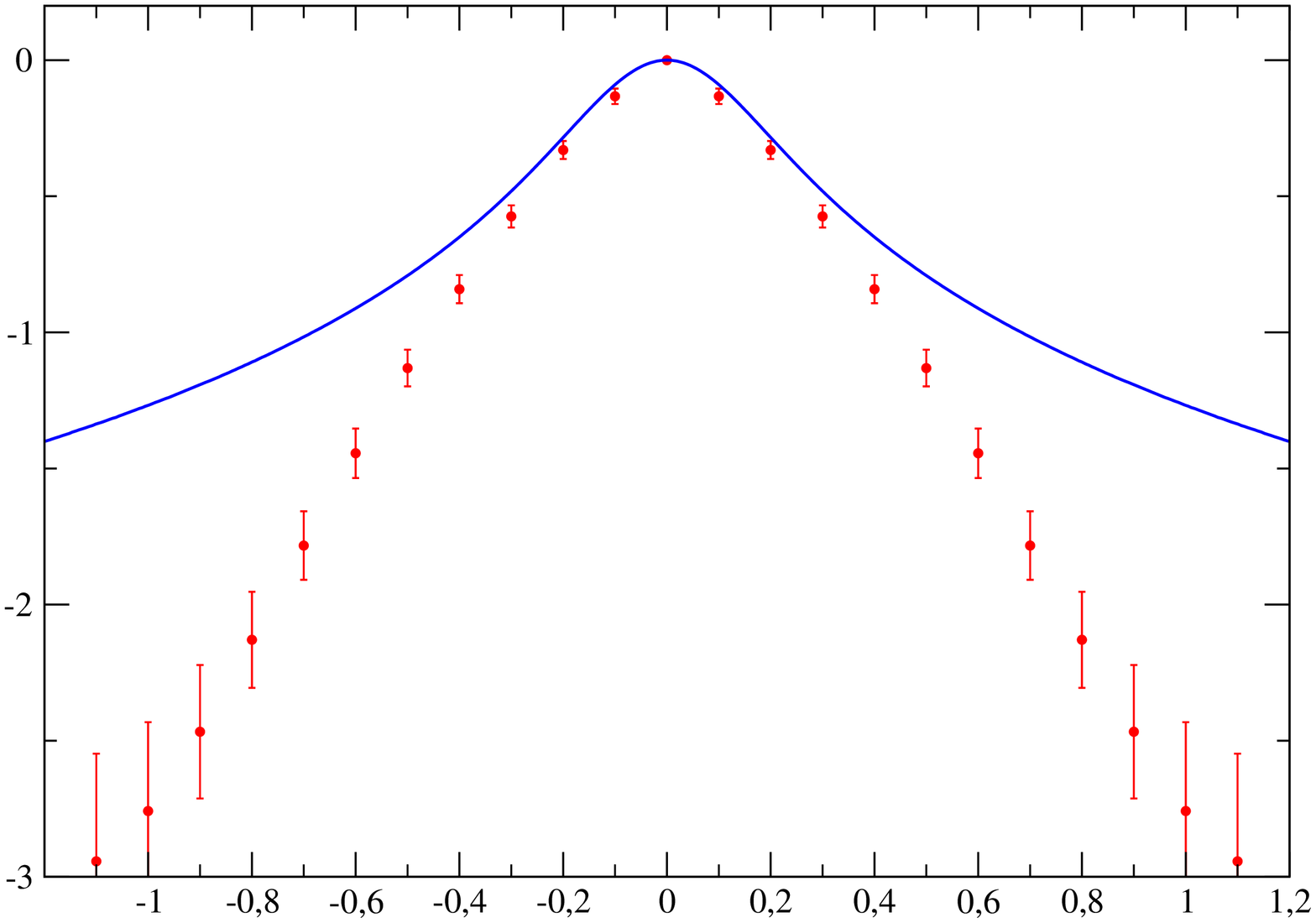} \\ [-1.0cm]
		  \hspace*{-0.8cm} 
		    \includegraphics[clip=true,angle=-90,scale=0.3]{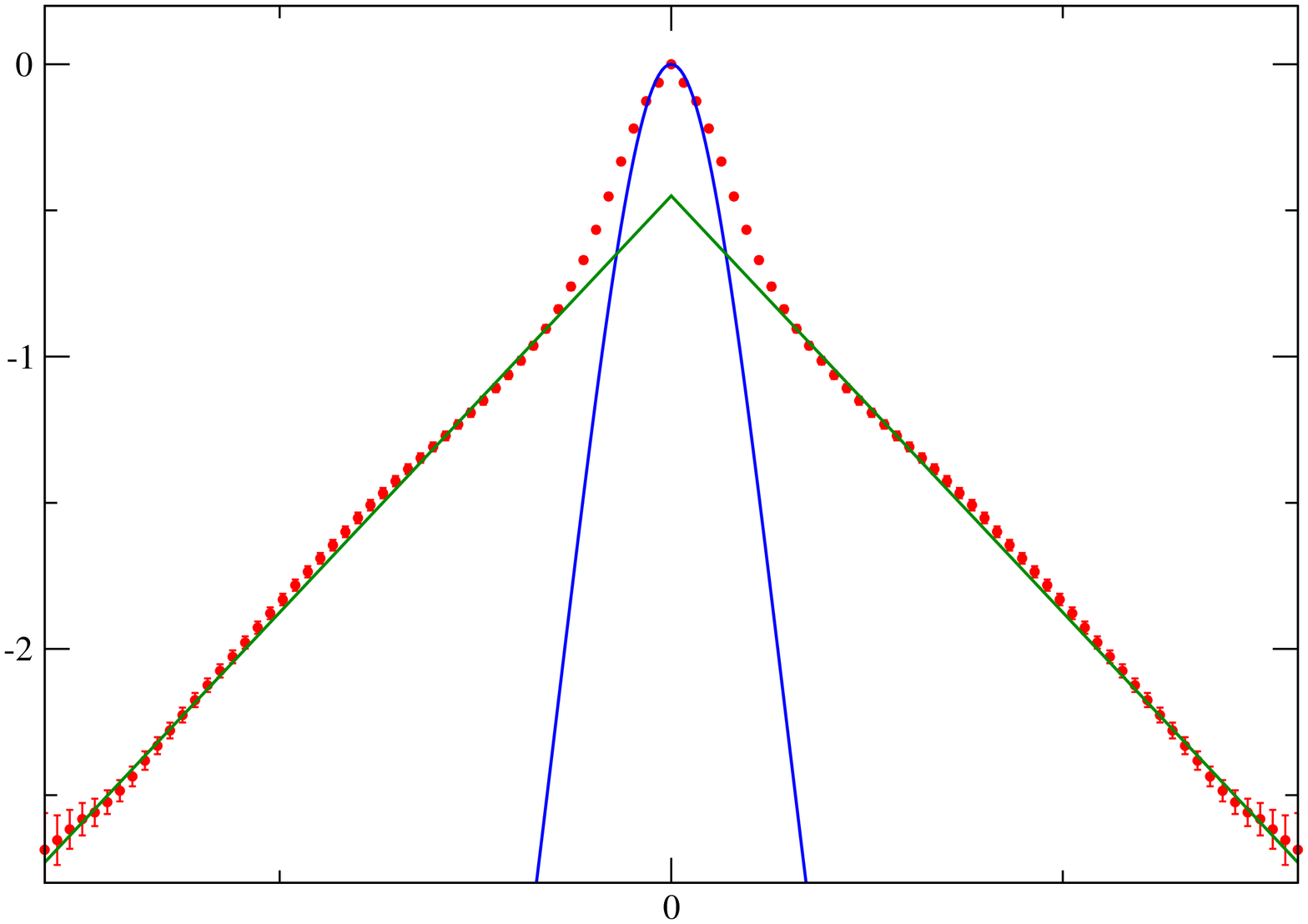} \\ [-1.0cm]
	    \hspace*{-0.8cm} 
	       \includegraphics[clip=true,angle=-90,scale=0.3]{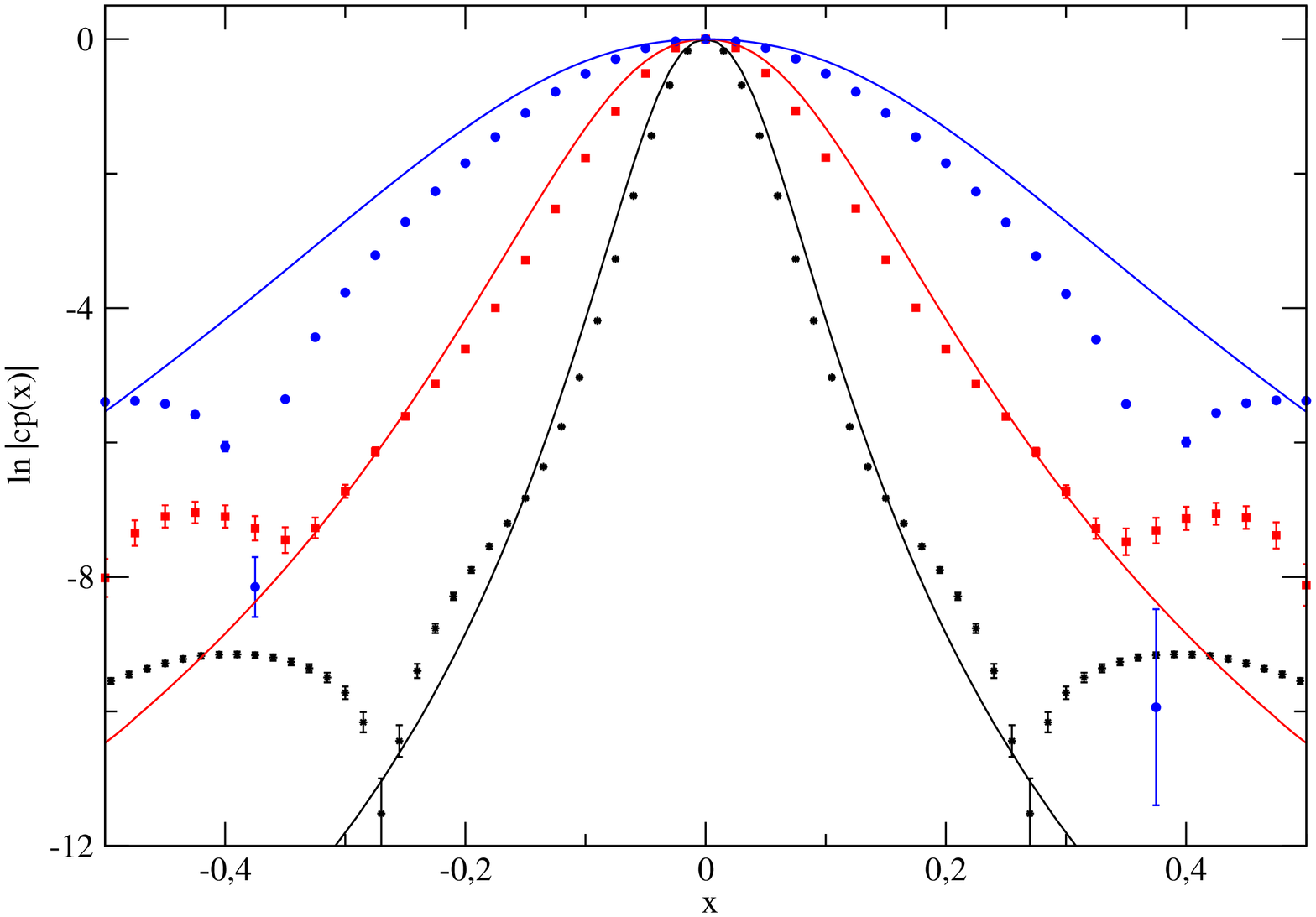}  \\
		         \end{tabular}	
	  \vspace*{-1.5cm}
	  \caption{ Correlation function of ${\bf E}^2 + {\bf B}^2$ for merons (top) and  instantons (middle) and of ${\bf E}\cdot {\bf B}$ for instantons (bottom).}
	   \label{fig:corr}
	 \vspace*{-.7cm}
		\end{figure}

If the effective theory contains the essential degrees of freedom, it will enable one to extract information about excited states from correlation functions: $\langle {\cal O}(t) {\cal O}(0) \rangle \sim \sum_n | \langle n | {\cal O} | \Omega \rangle |^2 e^{-E_nt}$. Thus, momentum projected correlation functions have been calculated of the following operators to obtain glueball masses with the indicated quantum numbers:
${\bf E}^2$ $ (0^+)$, ${\bf E}^2 \pm {\bf B}^2$ $(0^+)$, $\epsilon^{abc}E_i^a ({\bf E}^b {\bf B}^c)$ $ (1^+)$,  $\epsilon^{abc}B_i^a ({\bf E}^b {\bf B}^c)$ $ (1^-)$,  and $ E_i E_j -\frac{1}{3} {\bf E}^2 \delta_{i j} - B_i B_j +\frac{1}{3} {\bf B}^2 \delta_{i j} $ $ (2^+)$.  

The top graph in Fig.~\ref{fig:corr} demonstrates screening by the meron ensemble for the case of ${\bf E}^2 + {\bf B}^2$, where the solid curve denotes the correlation function for a single meron, which decays like $t^{-1}$ to within logarithmic corrections and the measurement for the ensemble generates a rapidly decaying exponential from which the $0^+$ mass may be measured. Similarly, the middle graph displays the emergence of a hadronic scale for an instanton ensemble, where the solid line shows the very compact correlation function for a single instanton and the ensemble measurements generate a much more slowly decaying exponential. 
Glueball masses extracted from instanton ensemble correlation functions
 agree qualitatively with lattice QCD\cite{Teper:1998kw}  within a factor of two.



The bottom graph shows the magnitude of the point-to-point correlation function  for the  topological charge density, ${\bf E \cdot  B}$, which has a positive short-range peak  close to the correlation function for a single instanton, playing the role of the positive delta-function contact term.  Beyond this peak, the correlation function  is negative, as expected from reflection positivity and the integral yields the physical values of $\chi$ shown in Table~\ref{tab:scaled}.


In summary, ensembles of regular gauge instantons and merons produce confinement and yield values of the action density, topological susceptibility, and glueball masses in qualitative agreement with lattice QCD.


\end{document}